\documentclass[conference]{IEEEtran}
\IEEEoverridecommandlockouts
\usepackage{cite}
\usepackage{amsmath,amssymb,amsfonts}
\usepackage{algorithmic}
\usepackage{graphicx}
\usepackage{textcomp}
\usepackage{xcolor}
\usepackage{caption}
\usepackage{subcaption}
\usepackage{booktabs}
\usepackage{url}
\usepackage{hyperref}

\usepackage{fancyhdr}
\fancypagestyle{firstpage}{
  \fancyhf{}                         
  
  \fancyfoot[C]{%
    \footnotesize
    \parbox{\textwidth}{\centering
      \textcopyright~2026 IEEE. Personal use of this material is permitted.
      Permission from IEEE must be obtained for all other uses, in any current or future media, including reprinting/republishing this material for advertising or promotional purposes, creating new collective works, for resale or redistribution to servers or lists, or reuse of any copyrighted component of this work in other works.\\[0.3em]
    }%
  }
}
\def\BibTeX{{\rm B\kern-.05em{\sc i\kern-.025em b}\kern-.08em
    T\kern-.1667em\lower.7ex\hbox{E}\kern-.125emX}}
\begin{document}
\bstctlcite{IEEEexample:BSTcontrol}
\title{Neural Compression of 360-Degree Equirectangular Videos using Quality Parameter Adaptation}

\author{\IEEEauthorblockN{Daichi Arai, Yuichi Kondo, Kyohei Unno, Yasuko Sugito, and Yuichi Kusakabe}
  \IEEEauthorblockA{\textit{Science \& Technology Research Laboratories}, \textit{NHK}, Tokyo, Japan\\
    \{arai.d-es, kondou.y-eg, unno.k-iw, sugitou.y-gy, kusakabe.y-ee\}@nhk.or.jp}}

\maketitle
\thispagestyle{firstpage}

\begin{abstract}
This study proposes a practical approach for compressing 360-degree equirectangular videos using pretrained neural video compression (NVC) models. Without requiring additional training or changes in the model architectures, the proposed method extends quantization parameter adaptation techniques from traditional video codecs to NVC, utilizing the spatially varying sampling density in equirectangular projections. We introduce latitude-based adaptive quality parameters through rate–distortion optimization for NVC. The proposed method utilizes vector bank interpolation for latent modulation, enabling flexible adaptation with arbitrary quality parameters and mitigating the limitations caused by rounding errors in the adaptive quantization parameters. Experimental results demonstrate that applying this method to the DCVC-RT framework yields BD-Rate savings of 5.2\% in terms of the weighted spherical peak signal-to-noise ratio for JVET class S1 test sequences, with only a 0.3\% increase in processing time.
\end{abstract}

\begin{IEEEkeywords}
360-degree video coding, equirectangular projection, neural video compression, quality parameter adaptation
\end{IEEEkeywords}

\section{Introduction}
360-degree video content has emerged as a key component in the field of immersive media. In contrast to traditional television, viewing devices for 360-degree videos, such as head-mounted displays, render only a portion of the omnidirectional scene, necessitating a substantially higher resolution \cite{BT2123} to deliver truly immersive experiences. Efficient encoding and decoding methods capable of handling the extremely high-resolution demands of 360-degree video are required to enable real-time broadcasting of such content.

Recent advancements in neural video compression (NVC) \cite{lu2019dvc, li2021deep, sheng2022temporal, 9897989, li2022hybrid, li2023neural, li2024neural, jia2025towards, zhang2025flavc, regensky2025nvc360} have demonstrated competitive performance in terms of both processing speed and compression efficiency compared with state-of-the-art video codecs. Notably, DCVC-RT \cite{jia2025towards} has achieved real-time operation on consumer-grade graphics processing units (GPUs), outperforming the encoding capabilities of next-generation codec reference software, such as ECM\footnote{\url{https://vcgit.hhi.fraunhofer.de/ecm/}}. However, applying NVC methods to 360-degree videos introduces certain challenges. Equirectangular projection, a standard format for omnidirectional videos, maps pixels from a spherical domain to a two-dimensional plane, resulting in increased spatial distortion at higher latitudes. Ideally, NVC models for 360-degree videos should be specifically designed and trained to compensate for this distortion using datasets comprising 360-degree videos. However, because 360-degree videos are often considered to be ancillary services, devices encounter challenges while supporting both standard compression and specialized models for 360-degree videos on the decoder side.

\begin{figure}
  \begin{subfigure}{\columnwidth}
    \begin{subfigure}{0.33\columnwidth}
    \centering
    \includegraphics[width=\columnwidth]{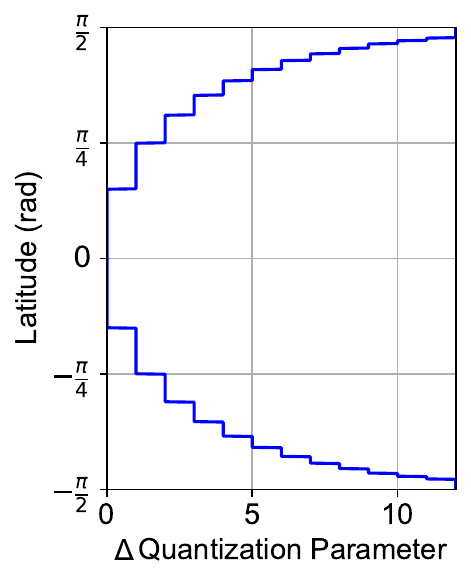}
    \end{subfigure}
    \begin{subfigure}{0.65\columnwidth}
    \centering
    \includegraphics[width=\columnwidth]{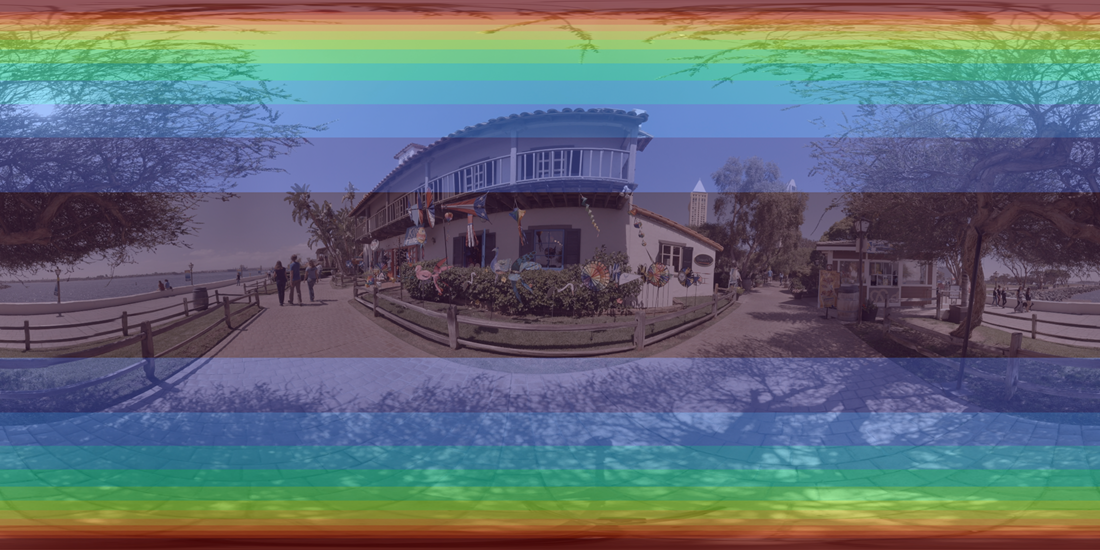}
    \vspace{4pt}
    \end{subfigure}
    \vspace{-5pt}
    \caption{Quantization parameter adaptation in video codecs}
  \end{subfigure}
  \begin{subfigure}{\columnwidth}
    \begin{subfigure}{0.33\columnwidth}
    \centering
    \includegraphics[width=\columnwidth]{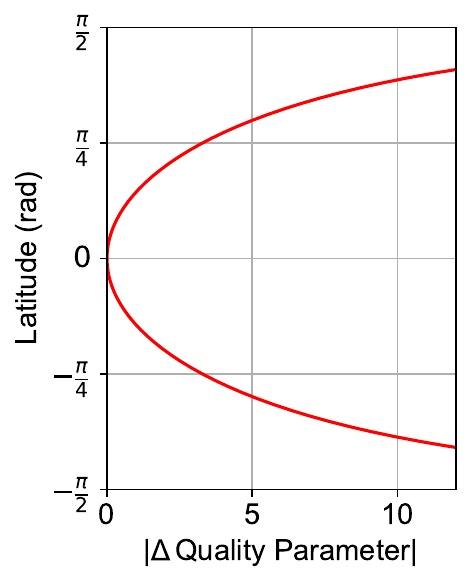}
    \end{subfigure}
    \begin{subfigure}{0.65\columnwidth}
    \centering
    \includegraphics[width=\columnwidth]{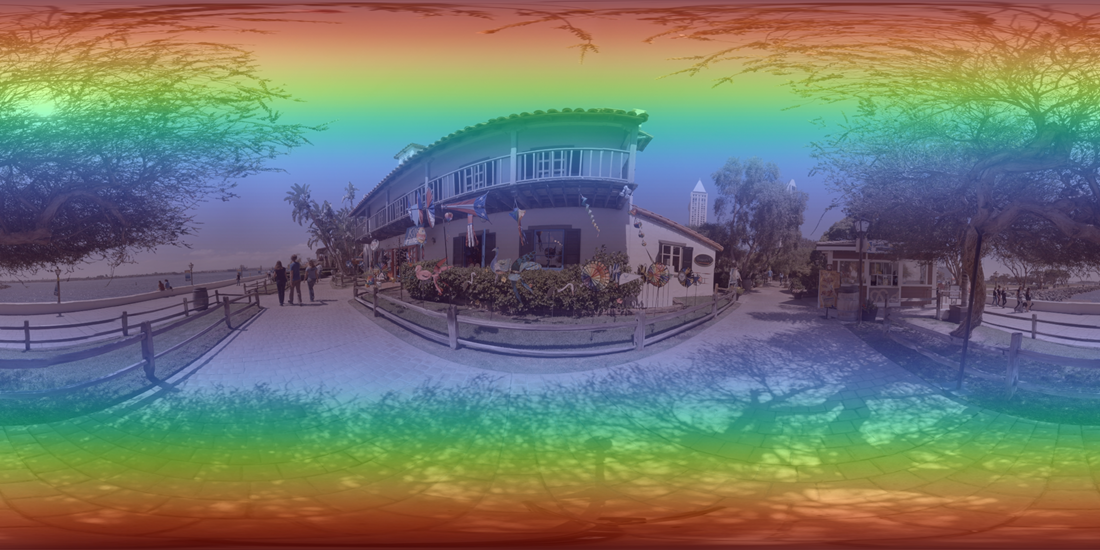}
    \vspace{4pt}
    \end{subfigure}
    \vspace{-5pt}
    \caption{Proposed quality parameter adaptation in DCVC-RT}
  \end{subfigure}
  \vspace{-15pt}
  \caption{Visualizations of quantization parameter adaptation in video codecs and proposed quality parameter adaptation in DCVC-RT \cite{jia2025towards} for 360-degree equirectangular videos.
  }
  \vspace{-15pt}
  \label{fig:1}
\end{figure}

We propose a practical neural compression technique for 360-degree equirectangular videos using existing pretrained models without requiring additional training or modifications to the NVC architecture. Specifically, we apply quantization parameter adaptation \cite{racape2017ahg8, hendry2017ahg8}, a concept originally explored in conventional video codecs, to the NVC realm. By leveraging the characteristics of the equirectangular format, where higher latitudes correspond to a larger spatial sampling density, the proposed method achieves an improved encoding performance, as evaluated using the weighted spherical peak signal-to-noise ratio (WS-PSNR) \cite{sun2017weighted} metric for 360-degree videos. Moreover, the proposed method utilizes a linear interpolation between two vectors within each vector bank for latent modulation, thereby enabling flexible adaptation with arbitrary quality parameters. Conventional video codecs that employ quantization parameter adaptation methods are constrained by rounding errors when calculating adaptive quantization parameters based on latitude, because of the discrete nature of these parameters. By contrast, NVC frameworks, such as DCVC-RT, facilitate latent modulation by training vector banks for each discrete quantization parameter, which enables the proposed method to perform linear interpolation between these vectors and achieve latent modulation as a function of latitude, as shown in Fig. \ref{fig:1}.

Section II reviews the related work on 360-degree video compression and NVC methods. Section III presents the preliminaries of the NVC framework. In Section IV, we introduce the latitude-based quality parameter adaptation. Section V describes the experimental settings and results. Finally, Section VI concludes the paper.

\section{Related Work}

\subsection{360-degree Video Compression}

Encoding methods for 360-degree videos have been extensively investigated \cite{8902161}. Strategies such as latitude-based quantization parameter adaptation \cite{racape2017ahg8, hendry2017ahg8} and padding techniques \cite{boyce2017ahg8} have been proposed to improve the encoding efficiency of equirectangular formats. Rate-distortion (RD) optimization formulations that consider the unique characteristics of 360-degree videos \cite{8379346, 8946760} and specialized prediction tools \cite{10290869, 8567927}, have also been introduced. 

\subsection{Neural Video Compression}

Various neural network-based video coding approaches, such as hybrid methods combining neural networks with conventional codecs \cite{jia2023deep, 10566387, ehrlich2024leveraging, 10849820}, have been widely studied. Among these, NVC methods \cite{lu2019dvc, li2021deep, sheng2022temporal, 9897989, li2022hybrid, li2023neural, li2024neural, jia2025towards, zhang2025flavc, regensky2025nvc360} excel in achieving superior compression efficiency, aiming to maximize performance at a given RD point. Traditional approaches typically train separate models, each optimized for a specific RD trade-off, thereby increasing training costs and limiting flexibility. By contrast, recent advancements, such as DCVC-FM \cite{li2024neural}, have proposed flexible models that can efficiently encode videos across a wide range of bitrates and quality levels using a single model. This approach is further enhanced by DCVC-RT \cite{jia2025towards}, which introduces implicit temporal modeling, eliminating explicit motion estimation networks and enabling the real-time coding of high-resolution videos across diverse practical bitrates.

\section{Neural Video Compression Framework}

\begin{figure}
    \centering
    \includegraphics[width=1.0\linewidth]{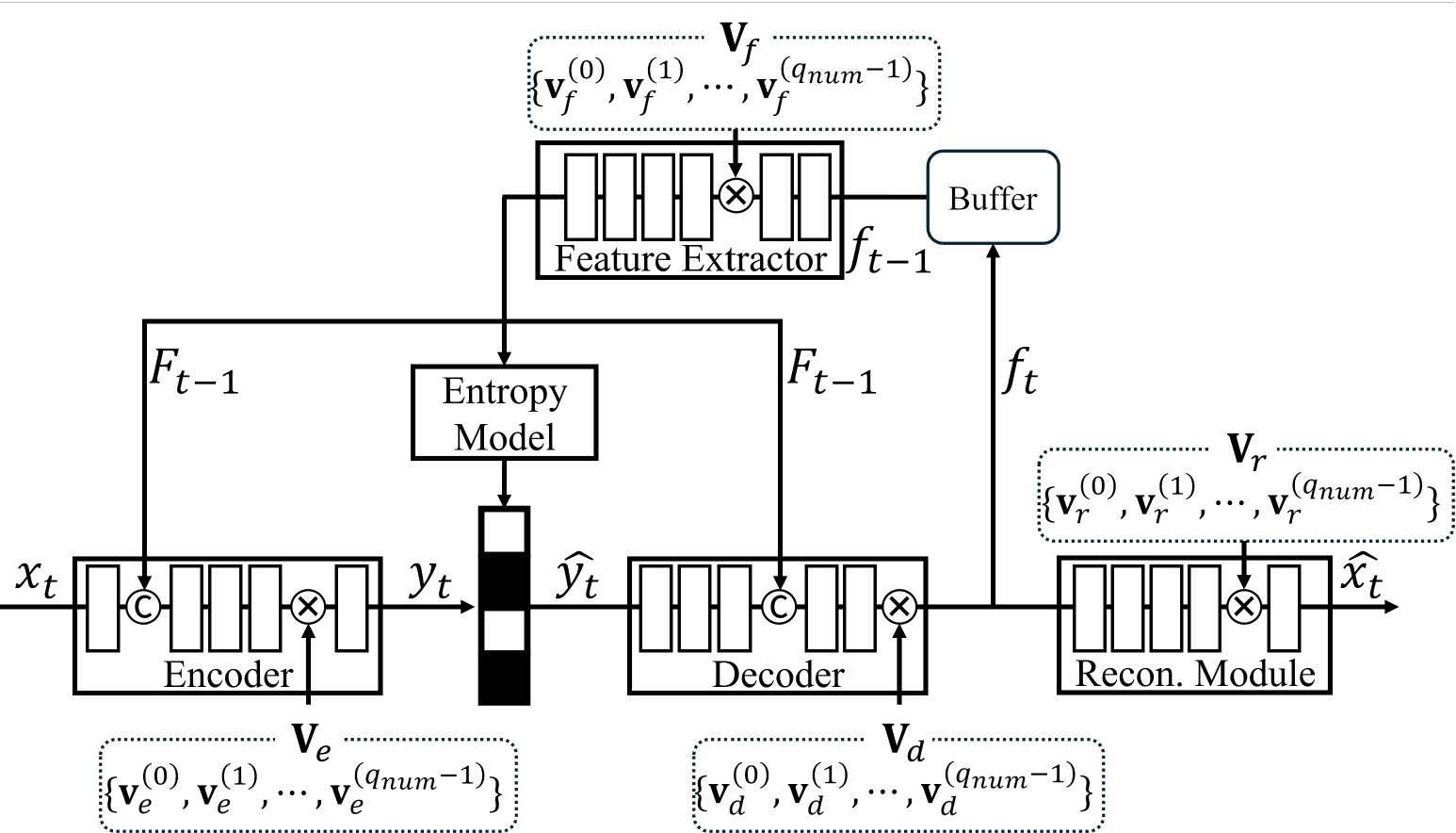}
    \caption{Simplified illustration of the DCVC-RT framework.}
    \label{fig:overview}
    \vspace{-10pt}
  \label{fig:2}
\end{figure}

Fig. \ref{fig:2} illustrates the DCVC-RT \cite{jia2025towards} framework, which supports a wide range of quality levels within a single model. At time $t$, the encoder takes the original frame $x_t$ as input and produces a latent $y_t$. The latent $y_t$ is then entropy coded by the entropy model and transmitted to the decoder. Subsequently, the decoder performs entropy decoding to obtain the decoded latent $\hat{y}_t$ and generates the reconstructed frame $\hat{x}_t$. The framework employs implicit spatial and temporal modeling by leveraging the latent $f_t$ to achieve efficient interframe coding. The output of the feature extractor at time $t-1$, $F_{t-1}$, is concatenated with the intermediate outputs of both the encoder and decoder to enhance coding efficiency. The framework also incorporates quality parameter-based latent modulation, achieved through channel-wise multiplication of the $\mathbf{v}^{(q)}_e$, $\mathbf{v}^{(q)}_d$, $\mathbf{v}^{(q)}_r$, and $\mathbf{v}^{(q)}_f$ vectors each corresponding to an integer quality parameter $q$ with the intermediate outputs of their respective network components, namely, the encoder, decoder, reconstruction module, and feature extractor.

An end-to-end RD optimization is performed using these vectors $\mathbf{v}^{(q)}_e$, $\mathbf{v}^{(q)}_d$, $\mathbf{v}^{(q)}_r$, and $\mathbf{v}^{(q)}_f$ in vector banks $\mathbf{V}_e$, $\mathbf{V}_d$, $\mathbf{V}_r$, and $\mathbf{V}_f$, respectively. Each model, in conjunction with its corresponding vectors, undergoes training to minimize the RD objective function:

\begin{equation}
J = R + \lambda D,
\end{equation}

\noindent where $J$ denotes the joint objective function that achieves an optimal RD point. During the training phase, the Lagrange multiplier $\lambda$ is determined within the range $[\lambda_{\text{min}}, \lambda_{\text{max}}]$ using an integer quality parameter $q$ randomly sampled from [0, $q_{\text{num}} - 1$], which is calculated as follows:

\begin{equation}
\lambda = e^{\ln(\lambda_{\text{min}}) + q \cdot \frac{\ln(\lambda_{\text{max}}) - \ln(\lambda_{\text{min}})}{q_{\text{num}}-1}},
\end{equation}

\noindent where $\lambda_{\text{min}}$, $\lambda_{\text{max}}$, and $q_{\text{num}}$ are constant parameters.

\section{Quality Parameter Adaptation}

\subsection{Calculation of Adaptive Quality Parameter}

\begin{figure}
  \begin{subfigure}{0.33\columnwidth}
    \centering
    \includegraphics[width=\columnwidth]{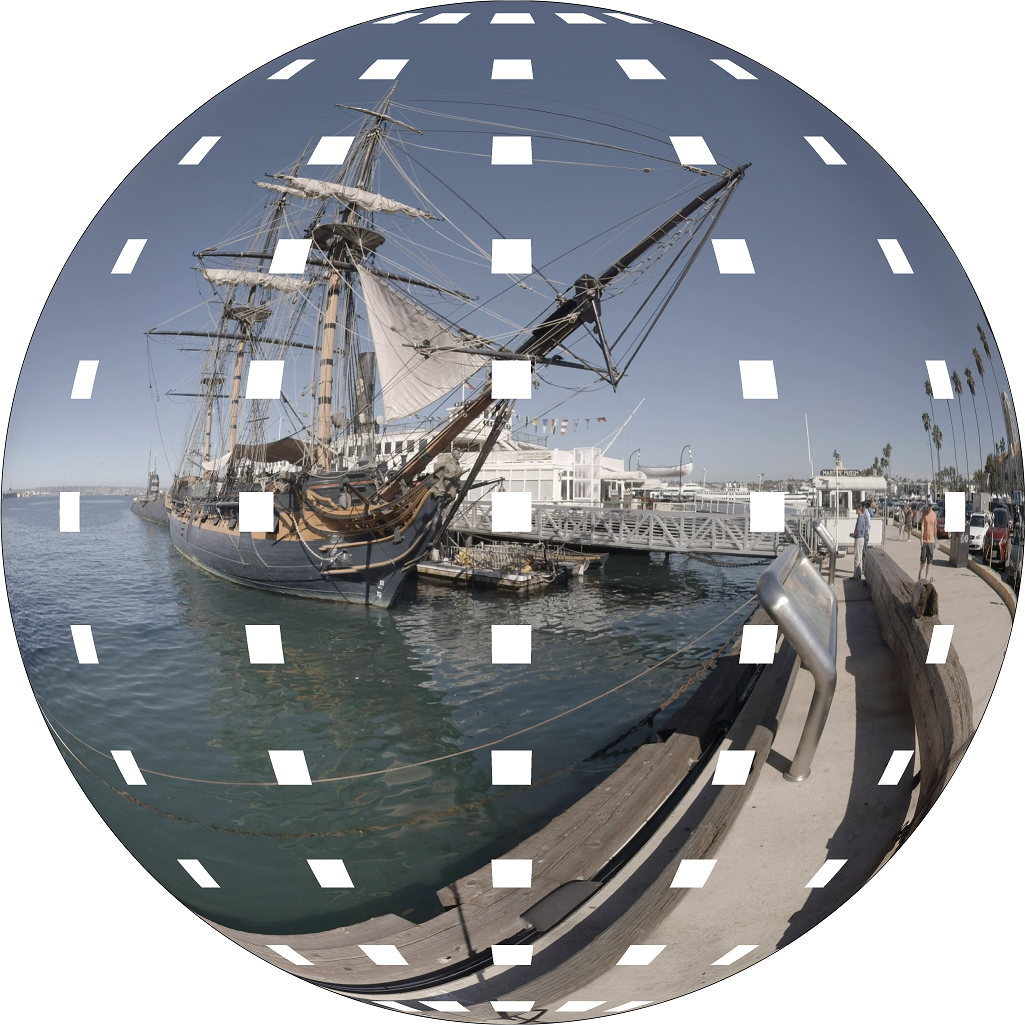}
    \caption{Spherical domain}
    \label{fig:sphere}
  \end{subfigure}
  \begin{subfigure}{0.66\columnwidth}
    \centering
    \includegraphics[width=\columnwidth]{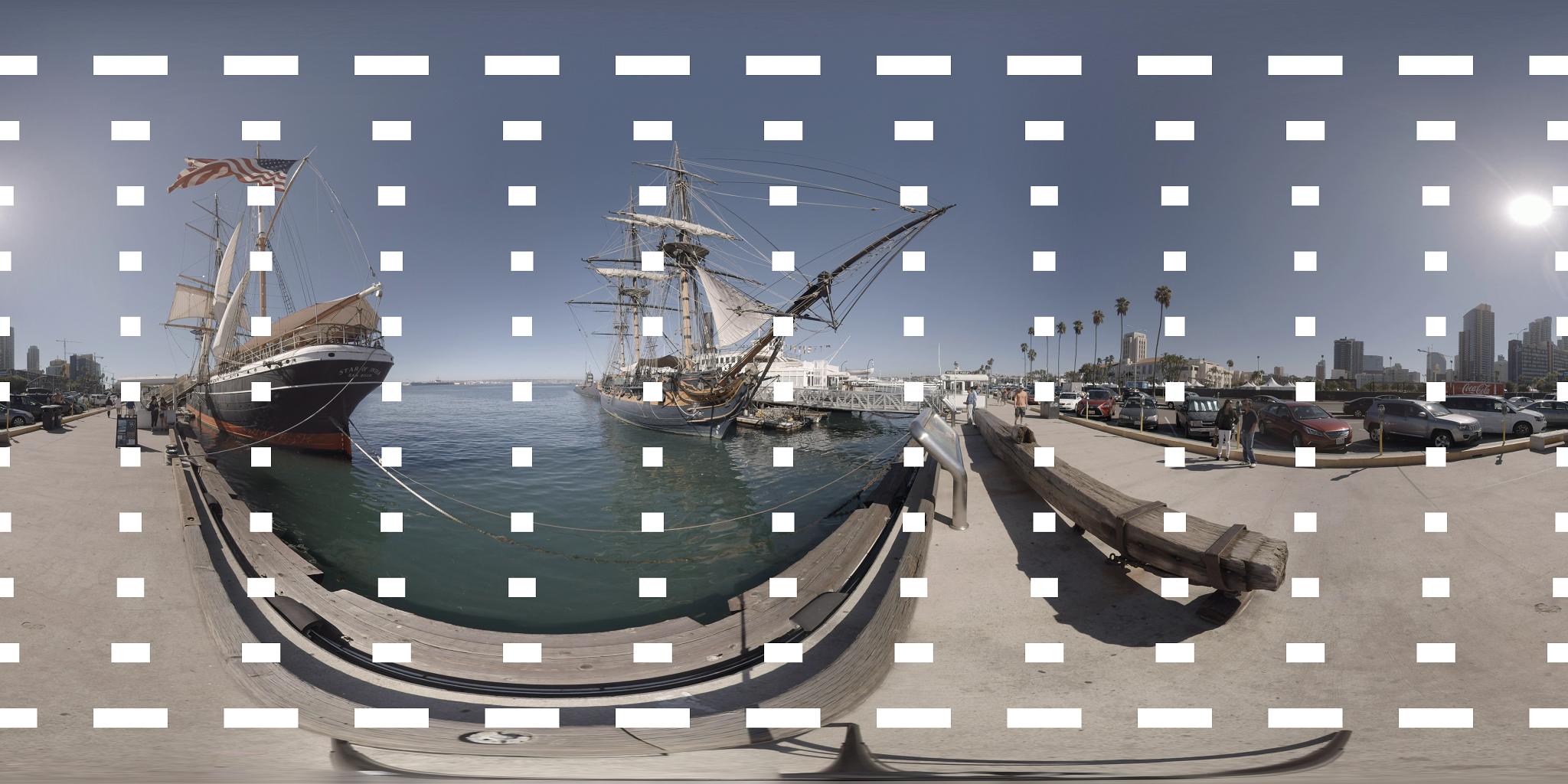}
    \caption{Equirectangular domain}
    \label{fig:equi}
  \end{subfigure}
  \caption{
  Conversion from spherical to equirectangular domain: equal-area rectangles become distorted and horizontally stretched with increasing latitude.
  }
\vspace{-10pt}
  \label{fig:3}
\end{figure}

We introduce quality parameter adaptation for NVC by extending the quantization parameter adaptation \cite{racape2017ahg8, hendry2017ahg8} explored in video codecs. 360-degree equirectangular videos exhibit an increased sampling density at higher latitudes, owing to the geometric transformation from the sphere to the plane. Let $\theta$ and $\varphi$ denote the longitude ($-\pi \leq \theta < \pi$) and latitude ($-\frac{\pi}{2} \leq \varphi \leq \frac{\pi}{2}$) on a sphere of radius $R$. In the spherical domain, the circumference of the horizontal circle at latitude $\varphi$ is given by $2 \pi R \cos \varphi$. By contrast, the equirectangular domain maintains a constant circumference across latitudes. This transformation stretches the circumference by a factor of $1/\cos \varphi$ when projected onto an equirectangular plane, as shown in Fig. \ref{fig:3}. Consequently, the area element $A_\varphi$ of the equirectangular projection can be expressed as

\begin{equation}
A_\varphi = \frac{A_0}{\cos \varphi},
\end{equation}

\noindent where \( A_0 \) denotes the area element at the equator (\( \varphi = 0 \)). 
To achieve uniform distortion distribution on the spherical domain, we require

\begin{equation}
D_\varphi = \frac{A_\varphi}{A_0}D_0 = \frac{D_0}{\cos \varphi},
\end{equation}

\noindent where \( D_\varphi \) and \( D_0 \) denote the distortion measures at latitude \( \varphi \) and the equator, respectively. Furthermore, the relationship between \(R\) and \(D\) in NVC is assumed to follow an exponential model \cite{sullivan2002rate}, where \(C\) and \(K\) are model parameters:

\begin{equation}
D = Ce^{-KR}.
\end{equation}

\noindent Solving (5) for \( R \) yields \(R = \frac{\ln C - \ln D}{K}\).
The Lagrange multiplier \(\lambda\) at the optimal RD point of (1) is given by \(\lambda = -\frac{\partial R}{\partial D} = \frac{1}{KD}\). 
From these relationships, the Lagrange multiplier \(\lambda_\varphi\) at latitude \( \varphi \) is obtained:

\begin{equation}
\lambda_\varphi = \frac{1}{KD_\varphi} = \frac{\cos\varphi}{KD_0} = \lambda_0 \cos\varphi,
\end{equation}

\noindent where \(\lambda_0 = \frac{1}{KD_0}\) is the base multiplier at the equator. By rearranging (2) with respect to \( q \), we obtain

\begin{equation}
q = \frac{(q_{\text{num}} - 1) \cdot (\ln(\lambda) - \ln(\lambda_{\min}))}{\ln(\lambda_{\max}) - \ln(\lambda_{\min})}.
\end{equation}

\noindent The adaptive quality parameter $q_\varphi$ is calculated as follows:

\begin{align}
q_\varphi &= q_0 + \Delta q_\varphi \\
\Delta q_\varphi &= q_\varphi - q_0\\
&=\frac{(q_{\text{num}} - 1) \cdot \ln(\cos \varphi)}{\ln(\lambda_{\max}) - \ln(\lambda_{\min})},
\end{align}

\noindent where $q_0$ denotes the base quality parameter and $\Delta q_\varphi$ represents the change of the adaptive quality parameter relative to the equator. As \( \varphi \) increases, $\Delta q_\varphi$ becomes more negative, resulting in a decrease in the quality at higher latitudes.

For the actual implementation, we apply $\tilde{q}_\varphi$, which is $q_\varphi$ minus the mean of $\Delta q_\varphi$ across all 
latitudes as follows:

\begin{gather}
\tilde{q}_\varphi = q_0 + \Delta q_\varphi - \bar{\Delta q} \\
\bar{\Delta q} = \frac{1}{\pi} \int_{-\pi/2}^{\pi/2} \Delta q_\varphi \, d\varphi 
= -\frac{(q_{\text{num}} - 1) \cdot \ln(2)}{\ln(\lambda_{\max}) - \ln(\lambda_{\min})},
\end{gather}

\noindent where $\bar{\Delta q}$ represents the mean of $\Delta q_\varphi$. This formulation aligns the mean of the adaptive quality parameter with $q_0$ for the entropy model, which uses a probability estimation module trained on the base quality parameter $q_0$.

\subsection{Vector Bank Interpolation for Latent Modulation}

In the DCVC-RT \cite{jia2025towards} framework, each vector bank contains vectors trained for distinct quality parameters. The conventional approach requires the quantization of the adaptive quality parameter $\tilde{q}_\varphi$ to select a single vector for latent modulation in each vector bank. Leveraging the arbitrary-rate capability of DCVC-RT, the proposed method addresses this limitation by employing a linear interpolation between the two nearest vectors. This technique enables the computation of vectors that match any arbitrary quality parameter $\tilde{q}_\varphi$ and can be applied during both the encoding and decoding processes. For instance, consider the vector bank of the encoder, denoted by $\mathbf{V}_e = \{\mathbf{v}^{(0)}_e, \mathbf{v}^{(1)}_e, \ldots\,, \mathbf{v}^{(q_{num}-1)}_e\}$. The interpolated vector $\mathbf{v}^{(\tilde{q}_\varphi)}_e$ is computed as follows:

\begin{equation}
\mathbf{v}^{(\tilde{q}_\varphi)}_e = (\lceil \tilde{q}_\varphi \rceil - \tilde{q}_\varphi) \cdot \mathbf{V}_e[\lfloor \tilde{q}_\varphi \rfloor] + (\tilde{q}_\varphi - \lfloor \tilde{q}_\varphi \rfloor) \cdot \mathbf{V}_e[\lceil \tilde{q}_\varphi \rceil],
\end{equation}

\noindent where \( \lfloor \cdot \rfloor \) and \( \lceil \cdot \rceil \) represent the floor and ceiling operations,
respectively. 
Although DCVC-RT is a frame-based architecture, the interpolated vector $\mathbf{v}^{(\tilde{q}_\varphi)}_e$ is adaptively adjusted according to the latitude derived from the vertical position of the latent and is channel-wise multiplied by the latent to enable latitude-based quality parameter adaptation.
Extending this linear interpolation approach to all vector banks $\mathbf{V}_e$, $\mathbf{V}_d$, $\mathbf{V}_r$, and $\mathbf{V}_f$ enables the proposed method to achieve flexible spatial and temporal quality adaptation for arbitrary quality parameters, thereby improving the encoding efficiency.

\begin{table*}[t]
\centering
\renewcommand{\arraystretch}{1.2}
\caption{Performance evaluation of the proposed method in terms of framerate and BD-Rate as measured by WS-PSNR.}
\begin{tabular}{lrr rrrrrrr}
\toprule
\textbf{Method} & \multicolumn{2}{c}{\textbf{Framerate (FPS)}} & \multicolumn{7}{c}{\textbf{BD-Rate (\%)}} \\
\cmidrule(lr){2-3}
\cmidrule(lr){4-10}
                & \textbf{Encoding} & \textbf{Decoding} & \textbf{SkateboardInLot} & \textbf{ChairLift} & \textbf{KiteFlite} & \textbf{Harbor} & \textbf{Trolley} & \textbf{GasLamp} & \textbf{Average} \\
\midrule
DCVC-RT Baseline     & 2.014 & 1.789 & 0.00  & 0.00   & 0.00   & 0.00  & 0.00   & 0.00   & 0.00 \\
w/o Interpolation    & 2.010 & 1.785 & \textbf{-4.27} & -11.37 & -6.48  & -2.70 & -2.34  & -3.31  & -5.08 \\
Proposed Method      & 2.008 & 1.784 & \textbf{-4.27} & \textbf{-11.43} & \textbf{-6.51}  & \textbf{-2.83} & \textbf{-2.49}  & \textbf{-3.65}  & \textbf{-5.20} \\
\bottomrule
\end{tabular}
\label{tab:dcvc_rt_variants}
\end{table*}

\begin{table*}[htbp]
\centering
\renewcommand{\arraystretch}{1.2}
\caption{BD-Rate (\%) comparison measured by WS-PSNR for quantization parameter adaptation applied to HEVC and VVC, and proposed quality parameter adaptation applied to DCVC-RT. HEVC results are obtained from \cite{racape2017ahg8}, and VVC and DCVC-RT results are based on our experiments. Note that each BD-Rate is calculated based on its reference. Lower is better.}
\begin{tabular}{@{}llrrrrrrr@{}}
\toprule
\textbf{Method} & \textbf{Reference} & \textbf{SkateboardInLot} & \textbf{ChairLift} & \textbf{KiteFlite} & \textbf{Harbor} & \textbf{Trolley} & \textbf{GasLamp} & \textbf{Average} \\
\midrule
HEVC (Random Access) & HM-16.15 & -9.10 & -8.80 & -2.87 & -1.16 & -1.39 & -1.19 & -4.09 \\
VVC (Random Access) & VVenC-slower-1.12.0 & -10.05 & -6.60 & -3.38 & -0.87 & -2.54 & -1.23 & -4.11 \\
VVC (Low-Delay P) & VVenC-slower-1.12.0 & -10.18 & -2.28 & -1.22 & 11.20 & 6.87 & 2.27 & 1.11 \\
DCVC-RT (Low-Delay P) & DCVC-RT & -4.27 & -11.43 & -6.51 & -2.83 & -2.49 & -3.65 & -5.20 \\
\bottomrule
\end{tabular}
\label{tab:qpa_effect}
\vspace{-10pt}
\end{table*}

\begin{figure*}
  \begin{subfigure}{0.666\columnwidth}
    \centering
    \includegraphics[width=\columnwidth]{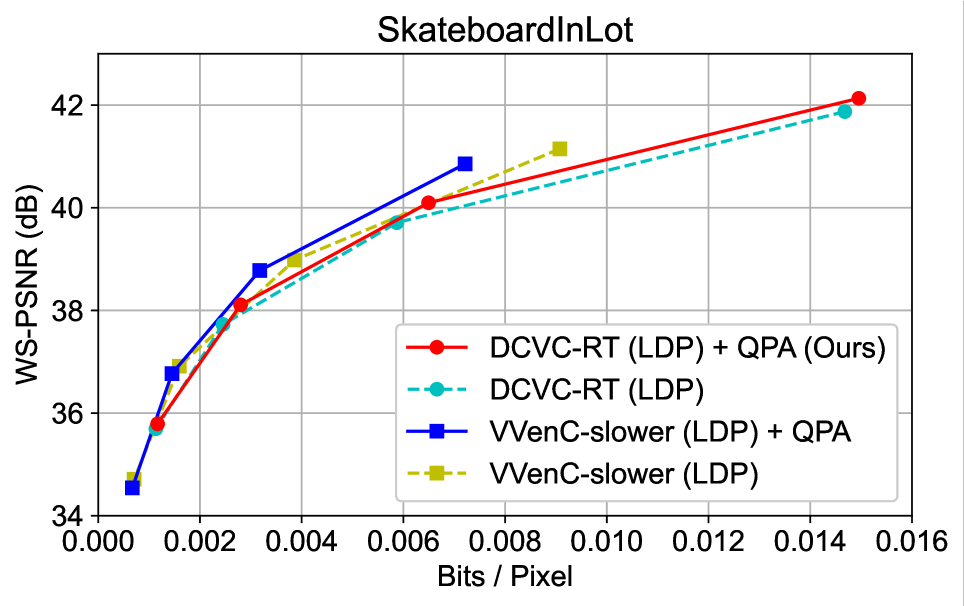}
    \label{fig:4a}
  \end{subfigure}
  \begin{subfigure}{0.666\columnwidth}
    \centering
    \includegraphics[width=\columnwidth]{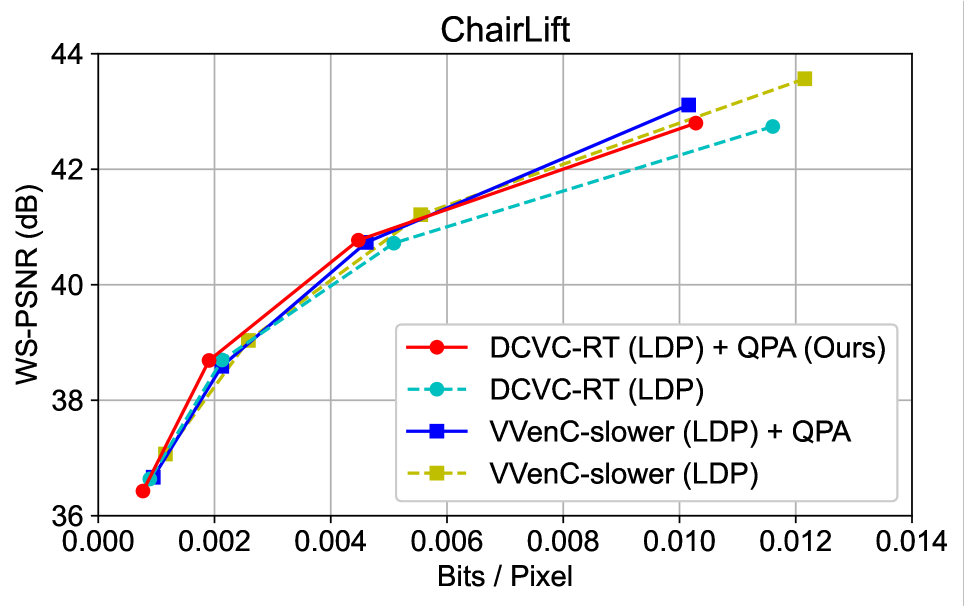}
    \label{fig:4b}
  \end{subfigure}
  \begin{subfigure}{0.666\columnwidth}
    \centering
    \includegraphics[width=\columnwidth]{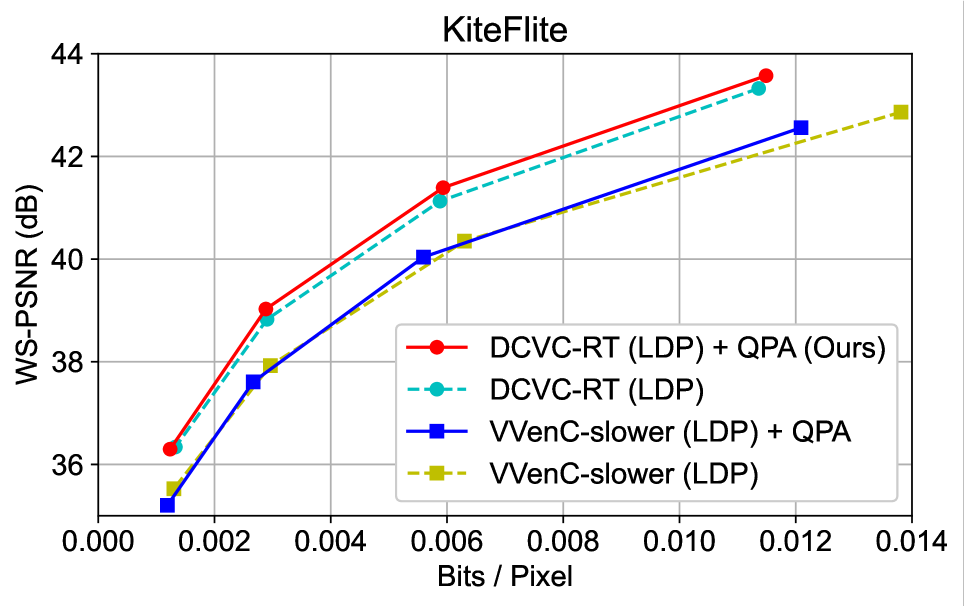}
    \label{fig:4c}
  \end{subfigure}
  \vspace{-10pt}
  \\
  \begin{subfigure}{0.666\columnwidth}
    \centering
    \includegraphics[width=\columnwidth]{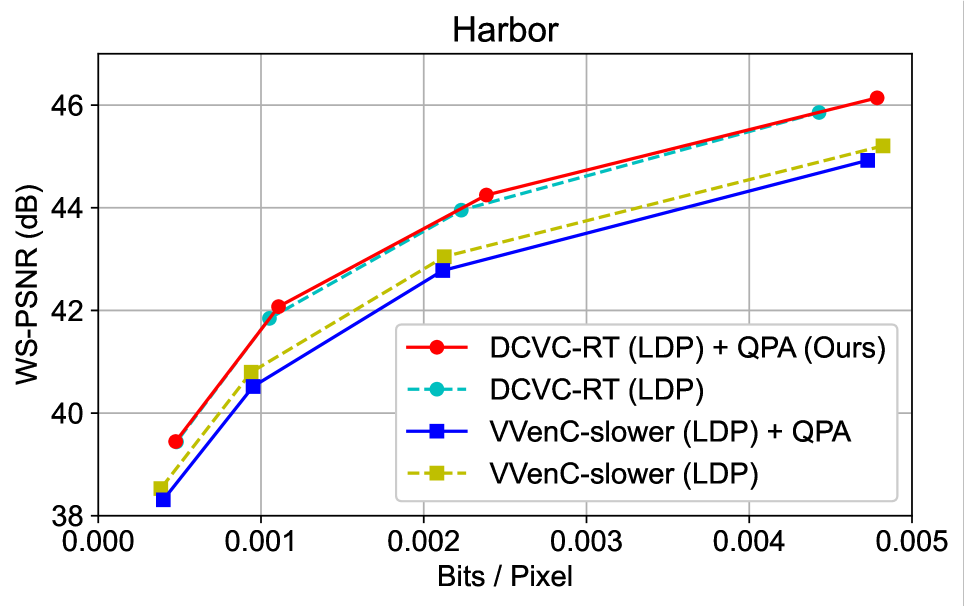}
    \label{fig:4d}
  \end{subfigure}
  \begin{subfigure}{0.666\columnwidth}
    \centering
    \includegraphics[width=\columnwidth]{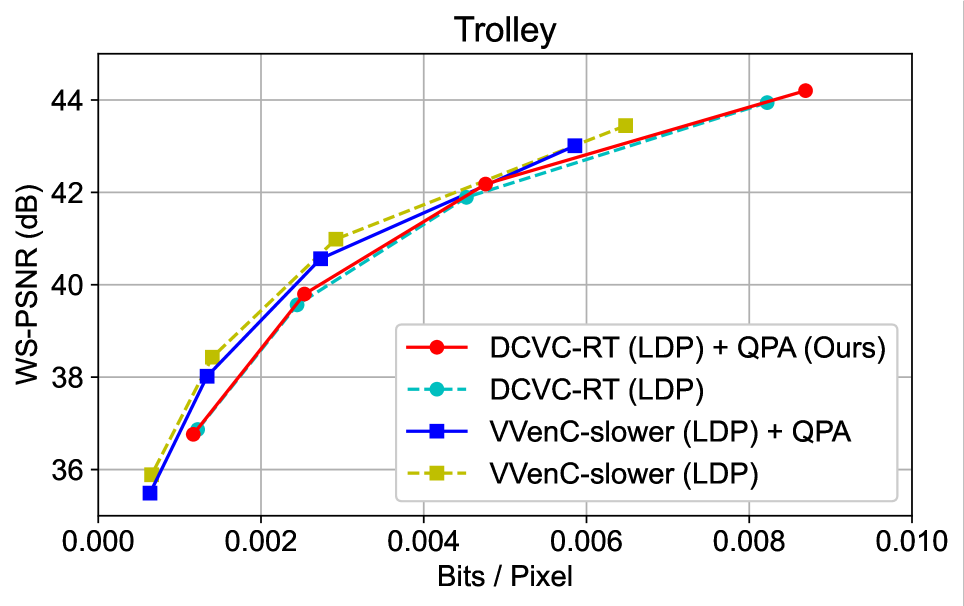}
    \label{fig:4e}
  \end{subfigure}
  \begin{subfigure}{0.666\columnwidth}
    \centering
    \includegraphics[width=\columnwidth]{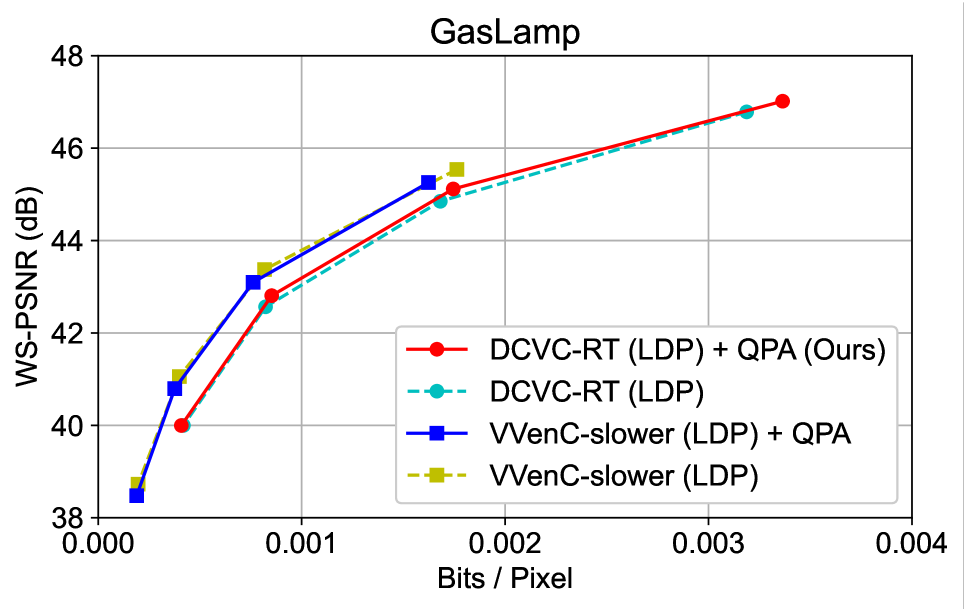}
    \label{fig:4f}
  \end{subfigure}
  \vspace{-15pt}
  \caption{RD curves for JVET class S1 test sequences. All frames were encoded with intra-period=\(-1\) under low-delay P (LDP) configurations. QPA denotes to quality parameter adaptation for DCVC-RT and quantization parameter adaptation for VVenC.}
  \vspace{-15pt}
  \label{fig:4}
\end{figure*}

\section{Experiments}

\subsection{Experimental Settings}

\noindent\textbf{Test Details}. We implemented the proposed method using DCVC-RT \cite{jia2025towards} with a pre-trained model. The DCVC-RT framework was configured in a low-delay P (LDP) configuration, using a quality parameter offset pattern of [0, 8, 0, 4, 0, 4, 0, 4] across eight frames. For the quality parameter adaptation, we used the same parameters as the pre-trained model: \( \lambda_{\max} = 768 \), \( \lambda_{\min} = 1 \), and \( q_{\text{num}} = 64 \). 
To enable comparisons with the latest video codec standards, we integrated quantization parameter adaptation into VVenC \cite{VVenC} version 1.12.0, which is a software encoder for versatile video coding (VVC) \cite{9503377}. In its slower configuration, VVenC achieves higher encoding efficiency than the reference software VTM\footnote{\url{https://vcgit.hhi.fraunhofer.de/jvet/VVCSoftware_VTM}}, while maintaining faster processing. We employed both the LDP and random access (RA) configurations to evaluate the encoding performance. In addition, we compared the results of quantization parameter adaptation \cite{racape2017ahg8} using HM\footnote{\url{https://vcgit.hhi.fraunhofer.de/jvet/HM}} version 16.15, a reference software for high efficiency video coding (HEVC) \cite{6316136}, in the RA configuration.

\noindent\textbf{Test Sequences}. For the evaluations, we utilized the JVET class S1 test sequences \cite{he2021jvet}, which comprise six uncompressed video sequences in equirectangular format, using the YUV color space with 4:2:0 chroma subsampling. The sequences had a pixel count of 8K and bit depths of both 8 and 10 bits. Each sequence was evaluated over 300 frames at the original resolution, for both DCVC-RT and VVenC experiments.

\noindent\textbf{Performance Evaluation}. To evaluate the video quality, we utilized WS-PSNR \cite{sun2017weighted} as the distortion metric for 360-degree videos, which is widely adopted in standard video coding activities. WS-PSNR extends PSNR by incorporating weightings at each pixel position based on the area covered in the spherical domain. WS-PSNR values were computed for the three YUV components using weights \((w_y, w_u, w_v) = (6, 1, 1)\). We employed BD-Rate \cite{BDBR} to evaluate RD performance. Furthermore, we investigated the computational efficiency of the proposed method by experimentally measuring the coding speed of DCVC-RT on a single NVIDIA V100 GPU with an Intel Xeon E5-2690 processor. We used the PyTorch implementation, instead of the native CUDA implementation.

\subsection{Experimental Results}

Table \ref{tab:dcvc_rt_variants} summarizes the performance of the proposed method in terms of encoding efficiency and coding speed. The proposed method achieved BD-Rate savings of 5.2\% in terms of WS-PSNR for the JVET class S1 sequences. Furthermore, Table \ref{tab:dcvc_rt_variants} summarizes the results of an ablation study without vector bank interpolation. In this case, we selected the vectors corresponding to the quantized \( \tilde{q}_\varphi \) for encoding and decoding. Although effectiveness varied across sequences, the proposed method consistently achieved enhanced coding performance across all the sequences examined. Moreover, the decrease in encoding and decoding frames per second was minimal, as the processing time increasing by less than 0.3\% compared with the baseline and by less than 0.1\% compared with the non-interpolation case.

Table \ref{tab:qpa_effect} lists the BD-Rate results for each video compression method when applying the quantization parameter adaptation or the proposed quality parameter adaptation, with the RD curves for the LDP configurations shown in Fig. \ref{fig:4}. Although the RA configurations consistently improve the BD-Rate results, the LDP configuration of VVenC exhibits performance degradation for some sequences. By contrast, the proposed method achieves stable improvements across all sequences under the LDP configuration. This suggests that in traditional coding schemes with LDP configurations lacking intra frames except at the beginning, quantization parameter adaptation may cumulatively degrade the reference frame quality in high-latitude regions, compromising the overall coding efficiency. However, the proposed adaptive quality adaptation method in DCVC-RT enables the simultaneous modulation of both spatial and temporal quality, resulting in stable performance gains in the LDP configuration.

\section{Conclusion}

This study proposed a practical approach for compressing 360-degree equirectangular videos using pretrained NVC models. We introduced adaptive quality parameters based on latitude to exploit the distortion caused by equirectangular projections, thereby enabling flexible adaptation through vector bank interpolation for latent modulation. Our experiments demonstrated a 5.2\% reduction in BD-Rate measured by WS-PSNR, with a 0.3\% increase in processing time. Further improvements in speed are anticipated with high-performance GPUs and native CUDA implementations. Future work includes developing of 360-degree video compression models that optimize encoding efficiency while maintaining backward compatibility with conventional videos, thereby enabling scalable and interoperable 360-degree video services.

\bibliographystyle{IEEEtran}
\bibliography{refs.bib}

\end{document}